\documentclass[12pt]{article}
\usepackage[english]{babel}
\usepackage{atbegshi,cite}
\usepackage{amsmath,amssymb,amsbsy,amstext, amsthm, simplewick}
\usepackage{hyperref}
\usepackage{graphicx}
\usepackage{amsfonts}
\usepackage[small]{caption}
\usepackage{upgreek}
\usepackage[titletoc]{appendix}
\usepackage[usenames,dvipsnames,table]{xcolor}
\usepackage{setspace}
\usepackage{color}
\newcommand{\newc}{\newcommand}
\newc{\fpi}{f_{\pi}}
\newc{\etap}{\eta^{\prime}}
\newc{\llll}{\langle\lambda\lambda\rangle}
\newc{\FFd}{F^a\tilde F^a}
\newc{\qbar}{{\overline q}}
\newc{\TR}{{\rm Tr}}
\newc{\Kahler}{K\"ahler }
\newc{\Zbb}{{\mathbb Z}}
\newc{\Rt}{{\mathbb R}^3}
\newc{\Rf}{{\mathbb R}^4}
\newc{\So}{{\mathbb S}^1}
\newc{\zt}{{\mathbb Z}_2}
\newc{\RtSo}{{\mathbb R}^3\times{\mathbb S}^1}
\newc{\scriminus}{{\cal I}^-}
\newc{\scriplus}{{\cal I}^+}
\newc{\mpl}{M_p}
\newc{\Ricci}{\mathcal{R}}
\newc{\bv}{\phi}
\newc{\hatr}{{\hat r}}
\newc{\calK}{K}
\newc{\calUi}{{\cal U}^{-1}}
\newc{\calE}{{\cal E}}
\newc{\calH}{{\cal H}}
\newc{\calL}{{\cal L}}
\newc{\calN}{{\cal N}}
\newc{\calM}{{\cal M}}
\newc{\calG}{{\cal G}}
\newc{\calO}{{\cal O}}
\newc{\calU}{{\cal U}}
\newc{\calUp}{{\cal U}^\prime}
\newc{\calUd}{{\cal U}^\dagger}
\newc{\calUpd}{{\cal U}^{\prime \dagger}}
\newc{\calX}{{\cal X}}
\newc{\calXp}{{\cal X}^\prime}
\newc{\calXd}{{\cal X}^\dagger}
\newc{\calXpd}{{\cal X}^{\prime \dagger}}
\newc{\calQ}{{\cal Q}}
\newc{\calOb}{{\cal O}^\dagger}
\newc{\hphi}{{\hat\phi}}
\newc{\llangle}{\langle\langle}
\newc{\rrangle}{\rangle\rangle}
\theoremstyle{plain}
\theoremstyle{plain} 
\theoremstyle{plain} 
\theoremstyle{plain}
\theoremstyle{plain}
\theoremstyle{plain}

\renewcommand{\title}[1]{{\Large\bf\flushleft{#1}}\vspace*{3ex}\\}
\renewcommand{\author}[2]{{\noindent\hspace*{2.5em}\large#1}
                     \footnote{Electronic mail: $\mathtt{#2}$}\\}

\begin{document}

\begin{titlepage}

\vskip 2.2cm

\begin{center}

{\large \bf  Breakdown of Field Theory in Near-Horizon Regions}
\vskip 1.4cm

{Tom Banks$^{(a),}$\footnote{tibanks@ucsc.edu},\,Patrick Draper$^{(b),}$\footnote{pdraper@illinois.edu},\;and Manthos Karydas$^{(b),}$\footnote{karydas2@illinois.edu}}
\\
\vskip 1cm
{ $^{(a)}$NHETC and Department of Physics \&\\Rutgers University, Piscataway, NJ 08854-8019}\\
{ $^{(b)}$Illinois Center for Advanced Studies of the Universe \&\\Department of Physics, University of Illinois, Urbana, IL 61801}\\
\vspace{0.3cm}
\vskip 4pt

\vskip 1.5cm

\begin{abstract}
We discuss  back-reaction in the semiclassical treatment of quantum fields near a black hole. When the state deviates significantly from Hartle-Hawking, simple energetic considerations of back-reaction  give rise to a characteristic radial distance scale $\sim (r_s^{2}G_N)^{1/D}$, below which some breakdown of effective field theory may occur.
\end{abstract}

\end{center}

\vskip 1.0 cm

\end{titlepage}
\setcounter{footnote}{0} 
\setcounter{page}{1}
\setcounter{section}{0} \setcounter{subsection}{0}
\setcounter{subsubsection}{0}
\setcounter{figure}{0}

\section{Introduction}
Effective field theory, perhaps the most useful theoretical tool in particle physics, is an imperfect description of quantum gravity, and its limitations can be interesting and subtle. 
In a recent paper~\cite{BP} it was argued, via quantum information considerations in field theory coupled to gravity, 
that if firewalls exist~\cite{AMPS} they must extend outside the black hole horizon to a distance of order $r_s^{2/D}$ in Planck units in $D$ dimensional spacetime.  In this note we show that the same length scale emerges when considering the gravitational back-reaction of the near-horizon field theory modes on the black hole metric, in states that deviate significantly from the Hartle-Hawking (HH) state.

The HH state itself exhibits small back-reaction on the geometry. The thermal properties of matter fields may be studied in the path integral on the smooth Euclidean Schwarzschild geometry. All UV divergences can be cut off and absorbed into the renormalization of terms in the gravitational effective action. The leading power-law divergences are absorbed into the cosmological constant, the Newton constant, and the GHY boundary term. When expressed in terms of the renormalized couplings, the total energy and entropy of the black hole receive only very small residual corrections. 

The fact that back-reaction is small in the HH state is rather less obvious in the Hilbert space picture. It has been discussed extensively in 't Hooft's brick wall model~\cite{tHooft:1984kcu}. 't Hooft proposed that the Bekenstein-Hawking entropy might be understood as a cutoff version of the entanglement entropy of quantum fields across the horizon. N\"aively, one finds that the  brick wall model of near horizon states is  inconsistent with the black hole metric at infinity and the HH vacuum, when the thermal entropy saturates the area law for the black hole~\cite{tHooft:1984kcu,Belgiorno:1995xc,Belgiorno:1995dt,Liberati:1996kt,Mukohyama:1998rf}. The basic reason is that the thermal energy in near horizon modes is also an order one fraction of the black hole mass. However, consistent with the general path integral reasoning, Susskind and Uglum~\cite{Susskind:1994sm} and Jacobson~\cite{Jacobson:1994iw} argued that one should again be able to see explicit cancellation in the Hilbert space picture between the ``bare gravitational" contributions to thermodynamic quantities and the divergent field theoretic contributions. This requires a covariant regularization scheme in which the brick wall can be removed, and it was demonstrated manifestly for the total entropy  in~\cite{Demers:1995dq}. For technical reasons described below, it is more difficult to sharply establish for the total energy, but nonetheless it must hold. Then, in either picture, accounting for the state-independent renormalizations of the gravitational couplings, the HH state has only small residual back-reaction on the horizon geometry.

On the other hand, Page's argument~\cite{Page:1993df,Page:1993up}, as interpreted by AMPS~\cite{AMPS}, suggests that it is interesting to think about black holes with fields that are very far from the HH state.\footnote{The distinction between the Unruh and HH states will not be relevant in our discussion.}  One can think of such states as resulting from the disentangling of pairs of Boulware modes near the horizon (and entangling the escaped Boulware particles with something else, for example, distant detectors, or earlier radiation). This disentangling picture can be viewed as producing a state which is a large excitation of HH. 

In this note, we estimate the gravitational back-reaction of large excitations of near-horizon modes away from the HH state. In contrast to  \cite{Demers:1995dq,Kim:1996bp} we leave 't Hooft's brick wall finite, but on a shorter distance scale than the covariant regulator. We show that this can be interpreted as a model  for a class of deviations from the HH state in which a controllable fraction of the near-horizon modes are removed from the thermal bath, modifying the thermal entropy. We can  see explicitly in this model how perturbative renormalization fails to completely cancel the ultraviolet sensitivity of the thermodynamic quantities. We find that when the deviation from HH is O(1), in the sense of an O(1) change in the thermal entropy of near-horizon modes, the energetic back-reaction implies that the QFT treatment breaks down below length scales of order $r_s^{2/D}$. This provides some complementary evidence for the findings of~\cite{BP}. More generally we interpret the result as indicating that generic excitations of the near-horizon QFT modes with wavelengths shorter than $r_s^{2/D}$ are not reliably described by QFT. This implies that the amount of black hole entropy that can be attributed to the entanglement entropy of quantum fields is at most $A^\frac{D-2}{2}$.

%For smaller deviations, we find a QFT cutoff that depends on $r_s$ and on the parameter controlling the magnitude of the deviation from HH.

Our analysis can be thought of as a generalization of the Cohen-Kaplan-Nelson (CKN) bound~\cite{ckn}. CKN considered gravitational backreaction on large excitations of the ordinary perturbative QFT vacuum. Hot boxes collapse to black holes, which are certainly out of field theoretic control, when the entropy increase is of order $A^{\frac{D-1}{D}}$. Here we replace ``perturbative QFT vacuum in flat space" by ``the HH state around a black hole" and ``heating the fields" by ``cooling some modes back to the Boulware state," then ask when the gravitational back reaction significantly changes the geometry.

Note added: after this work was complete, we were made aware of several related papers~\cite{Casher:1996ct,Sorkin:1996sr,Tuchin:1998fm,Marolf:2003bb,Anastopoulos:2014zqa,Brustein:2015sma,Srivastava:2020cdg,Ong:2023lbr}. These works estimate back-reaction effects on the classical horizon in a variety of different settings, including fluid modeling of  thermal atmosphere, modeling of the black hole interior, and analysis of near-horizon matter fluctuations, obtaining results consistent with those presented here by different means. In particular the work of Marolf in~\cite{Marolf:2003bb} (developing a Newtonian analysis of Sorkin~\cite{Sorkin:1996sr}) contains some arguments similar in spirit to ours. Marolf considers the backreaction effects of a local energy fluctuation and obtains in Eq. 3.7 a ``quantum width" of the horizon consistent with the scale obtained here. In this and other works the excitations are thought of as fluctuations in the thermal state. Our point of view, however, is somewhat different. The  Euclidean path integral is rendered completely finite by ordinary renormalization, to all orders in perturbation theory. Therefore, within the semiclassical expansion, one does not see back-reaction on the geometry from fluctuations in the HH state. We stay  within the semiclassical approximation, but consider back-reaction in large excitations of the HH state.

\section{The Regularized Brick Wall}
We consider $D=d+1$ dimensional static spherically symmetric solutions to vacuum Einstein equations with cosmological constant. 
The static coordinate metric takes the usual form
\begin{equation}
ds^{2}= -f(r)dt^2 + \frac{dr^2}{f(r)} + r^2 d\Omega_{d-1}^{2}\,.
\end{equation}
The horizon $r_{s}$ satisfies $f(r_{s})=0$.
The wave equation for a massive minimally coupled scalar field on this background is $$\frac{1}{\sqrt{-g}}\partial_{M}\left(\sqrt{-g}g^{MN}\partial_{N}\phi\right)-m^2\phi=0\,.$$We substitute $\phi= e^{-i\omega t}\, Y_{\ell,d}\,\phi(r)$, where $Y_{\ell,d}$ is the d-dimensional hyperspherical harmonic with angular momentum $\ell$, and obtain the radial mode equation
\begin{equation}
\label{radial_diff_on_curved_background}
\frac{\omega^2}{f(r)}\phi(r)+ \frac{1}{r^{d-1}}\partial_{r}(r^{d-1}f(r)\partial_{r}\phi(r)) - \frac{\ell (\ell +d-2)}{r^2}\phi(r)- m^{2}\phi(r)=0\,.
\end{equation}
In the leading order WKB approximation we set $\phi(r)= e^{i\Phi(r)}$ and drop $\Phi'$ and $\Phi''$ terms in Eq.~(\ref{radial_diff_on_curved_background}), retaining $\Phi'^2$. The result has the form  
\begin{align}
\Phi'(r)^{2}= \frac{1}{f(r)}\left(\frac{\omega^2}{f(r)}-\frac{\ell(\ell +D-3)}{r^2}-m^2\right).
\label{phiprimesqeq}
\end{align}
The WKB approximation is good for low frequencies in the near-horizon limit, where the omitted terms are suppressed relative to the $\Phi'(r)^{2}$ term by a redshift factor, and separately in the high frequency limit.

In the brick wall model we apply some boundary condition at coordinate radius $r_s+h$ and count oscillatory modes. The radial mode is oscillatory for $\Phi'(r)^{2}>0$. The outer solution to $\Phi'(r)^{2}=0$ determines some $r_{max}$\footnote{In the case of the de Sitter static patch, we have $\text{sgn}(f'(r_{s}))=-1$, $r_{s}=L$, $h<0$, and $r_{max}$ actually corresponds to a minimum  value of the radial coordinate. } as a function of the mode numbers $\omega$, $\ell$.  

We are interested in modes localized near the horizon. There are a large number of these states with frequency below the Hawking temperature in the thermal ensemble, $\omega r_s\lesssim 1$, because gravitational redshifting permits an unusual correspondence between ``short wavelength" and ``low frequency." We expand the phase near the horizon:
\begin{equation}
\Phi'(r)^{2}\simeq \frac{r_{s}^{2}\omega^2 - (r-r_s) f'(r_{s})[\ell(\ell +D-3)+m^2r_s^2]}{ r_{s}^{2}(r-r_s)^2 f'(r_{s})^2}\,
\end{equation}
where we have allowed the possibility that $m^2$ and $\ell^2$ may be of order $(r-r_s)^{-1}$. Factors of $|f'(r_s)|$ may be replaced by the horizon temperature $T_H$, $|f'(r_s)|=4\pi T_H\equiv {T}$. 
The positivity condition on $\Phi'(r)^{2}$ implies the following  $r_{max}$ and $\ell_{max}$:
\begin{equation}
r_{max}\simeq r_{s}+ \frac{ \omega^2 r_{s}^{2}}{ T\text{sgn}(f'(r_{s}))(\ell^2{+m^2r_s^2})}~~,~~\ell^{2}_{max}\simeq \frac{\omega^2 r_{s}^{2}}{T |h|}{-m^2r_s^2}\,.
\label{rmax}
\end{equation}
The approximations are self-consistent for $\ell$ or $m r_s$ large compared to one, and 
\begin{align}
    \rho_h\equiv \sqrt{r_s |h|}< \frac{\sqrt{r_s}\omega}{m\sqrt{T}}.
    \label{lhdef}
\end{align} %\MK{$\sqrt{r_{s}h}\to \sqrt{r_{s}|h|}$ in $\ell_{h}$}. 
$\rho_h$ is proportional to the proper distance from the brick wall to the horizon. Eq.~(\ref{lhdef}) imposes the condition that $\ell_{max}$ is positive, and it implies that we are working in the approximation that $\rho_h$ is the smallest physical distance scale in the problem. This  will be sufficient for our purposes.

The number of modes is below frequency $\omega$ is
\begin{equation}
\begin{split}
&N\simeq \frac{\text{sgn}(f'(r_{s}))}{\pi}\int^{\ell_{max}(\omega,h)}d\ell \frac{2\ell^{D-3}}{\Gamma (D-2)}\int_{r_{s}+h}^{r_{max}(\ell,\omega,h)}dr \,\Phi'(r)\\&
= \frac{2\omega}{\pi T}\int^{\ell_{max}(\omega,h)}d\ell \frac{2\ell^{D-3}}{\Gamma (D-2)}\left[ \text{tanh}^{-1}\left(\sqrt{\frac{\ell_{max}^2- \ell^2}{\ell_{max}^2{+m^2r_s^2}}}\right)-\sqrt{\frac{\ell_{max}^2- \ell^2}{\ell_{max}^2{+m^2r_s^2}}} \right].
\end{split}
\end{equation}

Now we consider the thermodynamics of one massless physical field and a collection of massive Pauli-Villars regulator fields~\cite{birrellanddavies}.  The regulators suffice to cancel formal divergences in mode sums, and in fact one can take the limit $h\rightarrow 0$ to achieve a complete covariant regularization of the HH state thermodynamics~\cite{Demers:1995dq}. We will  retain $h$, however, to parametrize the effects of short-distance deviations from the HH state. 

Let us introduce the small dimensionless ratios
\begin{align}
    z_i&\equiv m_i \sqrt{|h| r_s}\equiv m_i \rho_h,\nonumber\\
    y&\equiv |h|/r_s = (\rho_h/r_s)^2.
\end{align}
 For the massless field we have
\begin{align}
    N_0&\simeq a_{0}\left(\frac{\omega}{T}\right)^{D-1} \left(\frac{T r_s}{y}\right)^{\frac{D-2}{2}}\\&
   = a_{0} \left(\frac{\omega}{T}\right)^{D-1} (T r_{s})^{\frac{D-2}{2}}\left(\frac{r_{s}}{\rho_{h}}\right)^{D-2}\,,
\end{align}
where $a_{0}=\frac{1}{2^{D-3}(D-2)\Gamma (\frac{D-1}{2})\Gamma (\frac{D+1}{2})}$.

For the massive fields, 
\begin{align}\label{massiveNi}
N_{i}&=\left(\frac{\omega}{T}\right)^{D-1}(T r_{s})^{\frac{D-2}{2}} \left(\frac{r_{s}}{\rho_{h}}\right)^{D-2} f(b_{i},D)\,,\\\nonumber
f(b_{i},D)&=\frac{4 b_{i}^{D-2}}{\pi \Gamma (D-2)} \int_{0}^{1}dx (x^{D-3})\left[\text{tanh}^{-1}(b_{i}\sqrt{1- x^2})- b_{i}\sqrt{1- x^2}\right]\,,\\
    b_i^2&\equiv1-\frac{T  }{\omega^2 r_s}z_i^2.
\end{align}
The $x$ integral is finite. For frequencies such that Eq.~(\ref{lhdef}) is strongly satisfied, the $N_i$ can be expanded around $b_i\approx 1$ in the qualitative form
\begin{align}
\label{Niexpanded}
N_i&=\left(\frac{\omega}{T}\right)^{D-1}\left(r_s T\right)^{\frac{D-2}{2}} \left(\frac{r_s}{\rho_h}\right)^{D-2}\sum_{k=0}^{\infty} a_k c_i^k\nonumber\\
c_i^2&\equiv (1-b_i^2)=\frac{T}{\omega^2 r_{s}}z_{i}^2
\end{align}
where the $a_k$ are dimension-dependent numerical coefficients. This has the same form as $N_0$, modified by a power series in $c_i\sim z_i$. In even $D$, there are also $\log(c)$ terms. For example, in four dimensions one finds
\begin{align}
\sum_k a_k c_i^k &= \frac{2(\sqrt{1-c^2} \left(1+2 c^2\right)-3 c^2 \tanh ^{-1}\left(\sqrt{1-c^2}\right))}{3
   \pi }\nonumber\\
   &=\frac{2}{3 \pi }+\frac{c^2 \left(2 \log \left(\frac{c}{2}\right)+1\right)}{ \pi
   }-\frac{c^4}{4 \pi }+\dots
   \label{ackexp}
\end{align}

For low frequencies near the boundary of the inequality~(\ref{lhdef}), or equivalently $b_i\sim 0$, the expansion~(\ref{Niexpanded}) breaks down. One can instead expand in small $b_i$ and in this way find that $N_i\sim b_i^{D+1}$, indicating a strong suppression of the density of states in the infrared.

We can define a regularized state count by summing over the physical and PV fields,
\begin{align}
    N_{reg}\equiv \sum_i N_i.
\end{align}
Since the PV fields completely regulate the theory, $N_{reg}$   admits an ordinary Taylor series in $\rho_h$, as do all other thermodynamic quantities like the energy and entropy. Terms of order $\rho_h^{2-D}$ contributed by the individual fields cancel completely, and the collection of regulator masses is adjusted so the finite number of terms with negative powers of $\rho_{h}$  and logarithms also cancel. What remains from the  series in Eq.~(\ref{Niexpanded}) is $k= D-2$, which are terms independent of $\rho_h$, and  $k>D-2$, which are  terms proportional to positive powers of $\rho_h$. 

The PV masses are all $O(1)$ factors times a regulator scale. We write it as a length scale $\rho_c$:
\begin{align}
m_i\sim 1/\rho_c.
\end{align}
We can be more precise for specific choices of $D$. For the case of  $D=4$ with one massless physical scalar field, we require two anticommuting regulator fields of mass $1/\rho_c$, two commuting fields of mass $\sqrt{3}/\rho_c$, and one anticommuting field of mass $2/\rho_c$. The regulated state count is
\begin{align}
N^{D=4}_{reg}=\frac{ r_s^2
  }{ \pi  \rho_c^2}\left[ \frac{2 \omega }{T}\log\left(\frac{27}{16}\right)+ \frac{T}{4\omega^3 r_s^2}\left(\frac{\rho_h^4}{\rho_c^4}\right)+\dots\right].
\label{ND4reg}
\end{align}
for $\omega$ not too small. For small $\omega$ of order $\sqrt{T/r_s}(\rho_h/\rho_c)$, the small-$\rho_h$ expansion breaks down, and the actual  $N_{reg}$ (which can be obtained from (\ref{ackexp}) by summing over PV fields) is strongly suppressed.

For general $D$, the regulated state count has the small-$\rho_h$ expansion
\begin{align}
&N_{reg} = {\bar a}_{D-2} \left( \frac{r_s}{\rho_c}\right)^{D-2} \frac{\omega}{T}+\Delta N,\nonumber\\
&\Delta N \equiv \left( \frac{\omega}{T}\right)^{D-1} \left(r_s T\right)^{\frac{D-2}{2}} \left(\frac{r_s}{\rho_h}\right)^{D-2}\sum_{k=D-1}^{\infty} 
{\bar a}_k\left(\frac{T  }{r_s\omega^2}\right)^\frac{k}{2} \left(\frac{\rho_h}{\rho_c}\right)^k
\end{align}
for some modified  coefficients $\bar a_k$. 
The regularized density of states is\footnote{The literature often uses the notation $g(\omega)$ for the total state count below $\omega$, which we have called $N$.}
\begin{align}
    g(\omega) = dN_{reg}/d\omega\equiv g_0+\Delta g.
\end{align} 
 Let us estimate $g$  by retaining the leading terms in the $\rho_h$ series.  For $D=4$ we find
\begin{align}
    g(\omega)=\frac{2 r_s^2}{\pi \rho_c^2 T}\left(\log\left(\frac{27}{16}\right)-\frac{3
    }{8  (r_s T)^2  (\omega /T)^4}\left(\frac{\rho_h}{\rho_c}\right)^4+\dots\right).
    \label{gwD4}
\end{align}
For general $D$, it has the form
\begin{align}
g(\omega)\sim   r_s\left( \frac{r_s}{\rho_c}\right)^{D-2} X(\omega/T,r_s T) + r_s\left( \frac{r_s}{\rho_c}\right)^{D-2}\left(\frac{\rho_h}{ \rho_c}\right)^2Y(\omega/T,r_s T)+\dots
\end{align}
for some functions $X$, $Y$,.... (The $k=D-1$ term in $\Delta N$, if it exists, is independent of $\omega$. For the $D=4$ case we find the subleading terms start at order $(\rho_h/\rho_c)^4$.)

We will think of the first term, $g_0$, which is independent of $h$, as the  complete regularized density of states. It is dominated by angular $\ell$ modes of order $(r_s/\rho_c)$, rather than $\ell_{max}\sim (r_s/\rho_h)$, due to the Pauli-Villars cutoff. This will be important below. (It is evident, e.g., from the form of the individual $N_i$ in Eq.~\ref{massiveNi}, which scale as $\ell_{max}^{D-2}\sim (r_s/\rho_h)^{D-2}$. The leading term in $N_{reg}$ scales instead as $(r_s/\rho_c)^{D-2}$, reflecting the cancellation of contributions from wavelengths shorter than the regulator scale $1/\rho_c$.)

The second term, $\Delta g$, which is of order $h$, represents the ``nonthermal density of states." In thermodynamic sums we can interpret it as the density of states that are over- or under-populated with respect to the Bose-Einstein distribution corresponding to the HH state. We may think of $\Delta g$ as encoding a smooth modification of a fraction $(\rho_h/\rho_c)^2$ of the states  at every relevant energy scale, taking them to a nonthermal distribution. This is the sense in which the brick wall model can encode a one-parameter family of deviations from the HH state.

To address back-reaction we must estimate the $\rho_h$ dependent terms in the thermal energy,
\begin{align}
U = \int_0^\infty d \omega\,   g(\omega) \omega\left(\frac{1}{e^{\beta \omega}-1}\right)\equiv U_0+\Delta U.
\label{thermalU}
\end{align}
$U_0$ is the energy corresponding to the HH state and is independent of $h$.  $\Delta U$ most directly corresponds to a modification of the regulated density of states, relative to the regulated $h=0$ theory, but we could just as well think of it as a one-parameter family of excitations of the Hartle-Hawking state, parametrizing a deviation of some degrees of freedom out of the thermal state.

The total set of UV-sensitive contributions to the energy, including bare gravitational contributions, has the form:
\begin{align}
U_{tot}= [V\Lambda_0/G_{N,0}] + \left[\sum \frac{\omega}{2}\right] + [U_0+\Delta U] + \left[\frac{r_s^{D-3}F(r_s T)}{G_{N,0}}\right].
\label{totalU}
\end{align}
The first term contains the  bare cosmological constant, $\Lambda_0$, and the spatial volume $V$. The second term is the vacuum energy from the quantum fields. In leading-order WKB it has cutoff sensitivity $\sim  V/\rho_c^D $  which exactly cancels the perturbative renormalization of the $\Lambda_0$. The last term is the ``bare gravitational mass," with some function $F$ that depends on the black hole in question.

Had we written the entropy instead of the energy, with a similar decomposition $S_0+\Delta S$ coming from the quantum fields, the UV sensitivity in $S_0$ would exactly cancel with the perturbative renormalization of the Newton constant~\cite{Demers:1995dq,Kim:1996bp}. For $h=0$ the one-loop corrections that scale as ${\rho_c}^{2-D}$ merely replace $G_{N,0}\rightarrow G_{N,r}$ in the area law. 

The same is generally not quite true for the energy of the HH state. For example, for the Schwarzschild black hole in $D=4$ we easily obtain $U_0$ by inserting~(\ref{gwD4}) into~(\ref{thermalU}). We obtain
\begin{align}
   U_0 = \frac{r_s \log(27/16)}{48\pi\rho_c^2}
    \label{Uschw}
\end{align}
to be compared with the perturbative renormalization of the Newton constant appearing in $M=r/2G$,
\begin{align}
\frac{1}{G^{4D}_{N,r}}=\frac{1}{G^{4D}_{N,0}}+\frac{\log(27/16)}{6\pi\rho_c^2}.
\end{align}
Thus the bare mass renormalization is four times larger than the thermal energy in the field. This, however, is technical rather than profound. The Euclidean path integral on the smooth cigar geometry tells us that all UV sensitivity must be absorbed by counterterms in the HH state. It is difficult to establish a precise cancellation in the Hilbert space approach, but without too much difficulty it can be shown that NNLO WKB contributions to the vacuum energy $\sum\omega/2$ have subleading divergences of the correct form, cf.~(\ref{Uschw}), to fill the gap.\footnote{NLO WKB does not affect the phase $\Phi$, but at NNLO there is a correction that enters the density of states.} (This is the reason we included the vacuum energy and c.c. explicitly in the decomposition~(\ref{totalU}), instead of implicitly cancelling them.) 

In any case, it is a general property of the HH state that the ultraviolet contribution to $U_0$, scaling as $(r_s/\rho_c)^{D-2}$, cancels against various other contributions, including the vacuum energy and $G_{N,0}$.

For our purposes, the point is that $\Delta U$ is not absorbed by the counterterms, which are held fixed as we adjust $\rho_h$.\footnote{There could also be a contribution to the $\rho_h$-dependence in the energy from the vacuum energy, if we took the brick wall seriously as removing modes entirely. Since we are treating the modes as still present, just not in their thermal state, the vacuum energy is not sensitive to $\rho_h$.} This energy shift scales as
\begin{align}
    \Delta U   \sim  \left(\frac{1}{r_s}\right)\left(\frac{r_s}{\rho_c}\right)^{D-2}\left(\frac{\rho_h}{ \rho_c}\right)^n Z(\omega/T, r_s T, \beta \omega)\nonumber\\
    Z\equiv r_s\int^\infty d \omega\,   Y( \omega/T, r_s T) \left(\frac{r_s \omega}{e^{\beta \omega}-1}\right)
\end{align}
where $Z$ is a dimensionless function. We have omitted the lower limit of integration in $Z$ as a reminder that the $z\sim \rho_h/\rho_c$ expansion breaks down for  $\omega \rightarrow 0$, but the density of states is also very suppressed in this limit (cf. discussion after Eq.~(\ref{ackexp}), (\ref{ND4reg}).) The dominant contribution to the integral comes from $\beta \omega\sim 1$. The power $n$ might depend on the dimension. In $D=4$ we find $n=4$, and for $r_s\sim 1/T\sim \beta$,
\begin{align}
 \Delta U^{D=4} \sim \frac{r_s \rho_h^4}{\rho_c^6}.
 \end{align}

Finally we can estimate the effects of gravitational back-reaction. The quantum field computation only knows about the classical background geometry. Back-reaction is important when it changes the near-horizon geometry so much that the QFT computation breaks down.

Let us focus on the example of $D$-dimensional Schwarzschild. As discussed above, the regularized density of states for the near-horizon modes is dominated by angular modes 
\begin{align}
\ell_{dom}\sim r_s/\rho_c.
\end{align}
Inserting $\ell_{dom}$ into Eq.~(\ref{rmax}), we find that the dominant modes are localized in a radial shell of depth $(r_{max}-r_s)\sim r_s\left(\rho_c/r_s\right)^2$. Back-reaction on the perturbative QFT computations is therefore significant when there is an O(1) change in the geometry in this shell.  Typical near-horizon curvatures are of order $1/r_s^2$, so the back-reaction is important when the energy density  $\rho=T_0^0$ is of this order. The terms in $U_{tot}$ in (\ref{totalU}) correspond to the energy measured at infinity, which is related to the coordinate integral of $\rho$, so we find
\begin{align}
\frac{ G_{N,r} |\Delta U|}{(r_s^{D-2})(r_s\left(\rho_c/r_s\right)^2)}\sim \frac{1}{r_s^2}.
\label{criterion}
\end{align}
Eq.~(\ref{criterion}) is our general criterion for uncontrolled back-reaction. 

Physically the back-reaction effects may be thought of as follows. When $\Delta U$ is of this magnitude and positive, back-reaction from the state of the quantum field on the classical geometry increases the horizon radius by more than the size of the shell where the dominant modes are localized -- they ``collapse to a bigger black hole." In the particular state we are considering, $\Delta U$ is negative, behaving as a thin shell of negative energy near the horizon. In this case the most important influence on the geometry is to modify the near-horizon redshift factor,
\begin{align}
    g_{tt}(r\approx r_s)\rightarrow g_{tt}(r\approx r_s)+c \frac{G_{N,r}|\Delta U|}{r_s^{D-3}}
\end{align}
for positive constant $c$.
In the shell where the $\ell_{dom}$ modes are localized,  $r\sim r_s(1+(\rho_c/r_s)^2)$, the redshift factor is $g_{tt}\sim (\rho_c/r_s)^2$. So we see that the correction to the redshift is  significant when $\Delta U$ satisfies Eq.~(\ref{criterion}). An $O(1)$ increase in $g_{tt}$ lowers the local  temperature and increases
the Boltzmann suppression of every mode which is still in  thermal equilibrium. Equivalently, in the WKB analysis, many modes with $\omega\sim T_H$ and $\ell\sim \ell_{dom}$ suddenly find themselves with $(\Phi')^2<0$ in the corrected geometry (cf. Eq.~(\ref{phiprimesqeq})), removing them from the thermal bath.  This  reduces the QFT entropy and energy further, and in an uncontrolled way, in the sense that the correction-to-the-correction is of order the original correction. 

%The criterion is powerful because it is sensitive even to very small $\Delta r_s\ll r_s$. This is because the relevant QFT modes are sharply localized near the horizon.

For the $D$-dimensional Schwarzschild geometry, the correction to the energy  from the order-$\rho_h$ terms in the thermal energy (the ``firewall energy") is of the form
\begin{align}
\label{deltarsschwarz}
G_{N,r}  \Delta U  \sim r_s^{D-3} \left(\frac{\ell_p}{\rho_c}\right)^{D-2}\left(\frac{\rho_h}{\rho_c}\right)^n,
\end{align}
for some $n$, introducing the renormalized Planck length $\ell_p^{D-2}=G_{N,r}$. Inserting into Eq.~(\ref{criterion}), we find a breakdown of the QFT computation around a scale
\begin{align}
\rho_c\sim \rho_h^\frac{n}{D+n}r_s^\frac{2}{D+n}\ell_p^\frac{D-2}{D+n}. 
\label{ellcutoff1}
\end{align}
For a large excitation of the HH state, with $\rho_h\sim \rho_c$, there is an $O(1)$ deviation of the thermodynamic state of the static scalar modes: $\Delta g\sim g_0$, $\Delta U \sim U_0$, $\Delta S \sim S_0$, etc.  We may simply estimate the back-reaction by replacing $\Delta U$ with $U_0$ in the back-reaction criterion. For the Schwarzschild geometry, we find a breakdown of QFT at
\begin{align}
\rho_c\sim\rho_h\sim r_s^\frac{2}{D}\ell_p^\frac{D-2}{D}. 
\label{ellcutoff2}
\end{align}
This coincides with the scale found in~\cite{BP}.

\section{Discussion}

We interpret our result as meaning that if large excitations of a HH state are relevant in nature, one of two possibilities must hold. (1) They are still describable in semiclassical field theory, even very close to the horizon. Then the state must be quite different in nature from the toy example we studied, with a large modification to the entropy without a similar change in the energy. It is not clear if such a state exists consistent with entropy bounds in field theory.
(2) Semiclassical field theory simply breaks down on surprisingly long length scales near horizons.

%\subsection{Firewalls}
The authors of~\cite{AMPS} dubbed the failure of QFT near the horizon, revealed by  quantum information arguments, a {\it firewall}. An interpretation of~\cite{BP} is that if the firewall exists it must extend far outside the horizon. The authors of~\cite{Harlow:2013tf,Maldacena:2013xja,Akers:2022qdl} have argued that one can evade the conclusion of~\cite{AMPS} by noting that for most realistic systems, including black holes at time scales parametrically exceeding the Page time after collapse, the thermal equilibrium state is not really a typical state picked from the Gibbs ensemble.  Only the expectation values of certain coarse-grained operators have the same values for the true equilibrium state and the mathematical Gibbs state.  The AMPS argument proves only that typical pure states in the Gibbs ensemble have firewalls, and these authors suggest that the actual thermalized states approached in times polynomial in $r_s/\ell_p$ will be firewall-free.

Since back-reaction is a hydrodynamic property, depending only on the energy density, the distinction between ``typical" and ``actual" thermalized states is irrelevant to our calculation, and so cannot save us from the conclusion that something goes wrong with QFT in a large region outside the black hole horizon when the energy of the fields near the horizon differs significantly from the HH state. 

We emphasize that we are not claiming that unitarity of black hole evaporation implies a firewall. Among the alternatives described above, the most interesting is the possibility that  there is a breakdown of  quantum field theory at  distance scales from the horizon of order $r_s^{2/D}$ in Planck units. 
While we have done our calculation for a particular free field theory, with a particular cutoff, it seems likely that similar results hold in all perturbative field theories with a variety of choices of cutoff.  It would be interesting to attempt a more general CFT analysis of back-reaction, not assuming weak coupling.

Let us give a qualitative sketch how the ideas of~\cite{carlip,solo,tbwf,bfm,BZ} propose to resolve this problem.  According to those conjectures the density matrix of an empty diamond and an equilibrated black hole of the same size are identical.  The distinction between them has to do with how they are embedded in the Hilbert spaces of larger diamonds that contain them.  Localized objects in a diamond correspond to constrained states of the holographic system on the boundary of the diamond.  The number of constrained q-bits scales like $E_iR_{\diamond} $ for each localized object of energy $E_i$.  There are extra constraints that suppress degrees of freedom that couple the different localized objects.   This description is valid as long as $\sum E_i < R_{\diamond}^{d-3}$ in Planck units.  As we consider a low energy object approaching a black hole, we eventually reach a situation where $R_{\diamond}$ is no longer $\gg r_s$
and the picture of particles propagating around a black hole fails.  We will reserve detailed discussion of how this transition occurs to future work.

We have also mentioned the connection of our analysis with the CKN bound~\cite{ckn}. CKN made a proposal for how their cutoffs on the validity of QFT should actually be implemented in computations, resulting in mild effects on precision observables scaling as a fractional power of $G_N$~\cite{ckn,ck}. In~\cite{Banks:2019arz,Blinov:2021fzl} it was shown that the bound could be realized in a rather more gentle way as a depletion of the field theoretic density of states, such that the impact on precision observables is always of order $G_N$, which is negligible. A notable byproduct is that either interpretation of the bound eliminates the radiative fine-tuning problem of the cosmological constant. Perhaps a similar depletion interpretation could be assigned to the cutoffs~(\ref{ellcutoff1}), (\ref{ellcutoff2}). Curiously, while the CKN bound suggests that in a fixed region in a flat background the number of degrees of freedom that are well-modeled by QFT scales as $A^\frac{D-1}{D}$, the back-reaction analysis here suggests a near-horizon  entropy well-modeled by QFT that is bounded by $A^\frac{D-2}{D}$.

Implications for finite causal horizons and cosmological horizons, as well as alternative conclusions in the Holographic Space-time models of quantum gravity, will be discussed elsewhere.

~\\

\section*{Acknowledgements} 
The work of PD and MK is supported by the U.S.
Department of Energy, Office of Science, Office of High Energy Physics under award number DE-SC0015655. The work of TB is supported by the U.S.
Department of Energy, Office of Science, Office of High Energy Physics under award number DE-SC0010008. Rutgers Project 833012.

~\\

\bibliography{biblio}{}
\bibliographystyle{utphys}

\end{document}